\documentclass[12pt,a4paper]{article}
\usepackage{a4wide}
\usepackage{amsmath}
\usepackage{amssymb}
\usepackage{amsfonts}
\usepackage{epsfig}
\usepackage{subfigure}
\usepackage{exscale}
\usepackage{float}
\usepackage{bbm}
\usepackage[numbers,sort&compress]{natbib}

\newcommand{\p}{\partial}

\makeatletter
\@addtoreset{equation}{section}
\makeatother

\title{Constraint Effective Potential of the Staggered Magnetization in an 
Antiferromagnet}

\author{U.\ Gerber$^a$, C.\ P.\ Hofmann$^b$, F.-J.\ Jiang$^a$, M.\ Nyfeler$^a$,
U.-J.\ Wiese$^{a,c}$
\\ \\
$^a$ Center for Research and Education in Fundamental Physics \\
Institute for Theoretical Physics, Bern University \\
Sidlerstrasse 5, CH-3012 Bern, Switzerland
\\ \\
$^b$ Facultad de Ciencias, Universidad de Colima \\
Bernal D\'iaz del Castillo 340, Colima C.P.\ 28045, Mexico
\\ \\
$^c$ Institute for Theoretical Physics, ETH Z\"urich \\
Schafmattstrasse 32, CH-8093 Z\"urich, Switzerland \\ \\}

\begin{document} 

\maketitle

\begin{abstract} \normalsize

We employ an improved estimator to calculate the constraint effective potential
of the staggered magnetization in the spin $\tfrac{1}{2}$ quantum Heisenberg 
model using a loop-cluster algorithm. The first and second moment of the 
probability distribution of the staggered magnetization are in excellent 
agreement with the predictions of the systematic low-energy magnon effective 
field theory. We also compare the Monte Carlo data with the universal shape of 
the constraint effective potential of the staggered magnetization and study its
approach to the convex effective potential in the infinite volume limit. In 
this way the higher-order low-energy parameter $k_0$ is determined from a fit
to the numerical data.

\end{abstract}

\newpage

\section{Introduction}

Systematic low-energy effective field theory is a very powerful tool for
investigating the low-energy physics of Goldstone bosons. In particle physics 
this tool has been used to describe the low-energy physics of pions 
\cite{Wei79}. Pions arise as the pseudo-Nambu-Goldstone bosons of the 
spontaneously broken $SU(2)_L \times SU(2)_R$ chiral symmetry of QCD. Chiral
perturbation theory provides a systematic momentum expansion of the low-energy
physics of pions \cite{Gas85}. Based on symmetry considerations, the effective
theory makes detailed quantitative predictions, which depend on the values of
some a priori unknown low-energy parameters, such as the pion decay constant 
and the chiral order parameter. Systematic effective field theories have also 
been developed for quantum field theories and classical spin systems with a
spontaneously broken $O(N)$ symmetry \cite{Has90}. Effective field theories 
with an $O(3)$ symmetry have been used to describe the low-energy physics
of magnons --- the Goldstone bosons of quantum antiferromagnetism 
\cite{Cha89,Neu89,Fis89,Has91}. In this case, the relevant low-energy 
parameters are the spin stiffness $\rho_s$, the spinwave velocity $c$, and the
staggered magnetization density ${\cal M}_s$. In particular, the effective 
theory makes detailed predictions for the finite-size and finite-temperature 
effects of systems in large volumes at low temperatures. These effects have 
been worked out by Hasenfratz and Niedermayer even at the 2-loop level 
\cite{Has93}. By comparing the effective field theory predictions with Monte 
Carlo data obtained with an efficient loop-cluster algorithm \cite{Eve93}, the 
low-energy parameters $\rho_s$, $c$, and ${\cal M}_s$ have been determined with
high accuracy for the spin $\tfrac{1}{2}$ quantum Heisenberg model, both on the
square \cite{Wie94,Bea96} and on the honeycomb lattice \cite{Jia08}, as well as
for the $J$-$Q$ model with 2- and 4-spin interactions on the square lattice 
\cite{Jia08a}. Similar methods have been applied to the 4-d $O(4)$-symmetric 
quantum field theory describing the Higgs sector of the standard model 
\cite{Has90a}. By comparing with effective field theory predictions for the 
susceptibility of the order parameter, both the vacuum expectation value of the
Higgs field and the Higgs boson mass have been determined from Monte Carlo data
for the 4-d lattice $O(4)$ model \cite{Goe91b,Goe93} obtained with the Wolff 
cluster algorithm \cite{Wol89}. The systematic low-energy effective field 
theory also predicts the shape of the constraint effective potential of the 
order parameter, which has been worked out in great detail by G\"ockeler and 
Leutwyler \cite{Goe91,Goe91a}. Their predictions for the shape of the 
constraint effective potential have been tested against Monte Carlo simulations
of classical 3-d $O(3)$ and 4-d $O(4)$ lattice models \cite{Dim91}.

In this paper, we develop an improved estimator for the probability 
distribution of the staggered magnetization in order to extract the constraint
effective potential from loop-cluster simulations of the spin $\tfrac{1}{2}$ 
quantum Heisenberg model on the square lattice. The first moments of the 
probability distribution are in excellent agreement with the predictions of the
magnon effective theory. The Monte Carlo data approach the universal shape of 
the effective potential in the infinite volume limit, which is known to be a 
convex function \cite{Rai86}. Our study confirms that the effective field 
theory provides quantitative predictions for a wide class of low-energy
observables, which are exact, order by order in a systematic momentum 
expansion.

The rest of this paper is organized as follows. Section 2 summarizes the 
predictions of the low-energy magnon effective field theory that are relevant 
for our study, and section 3 discusses the determination of the low-energy 
parameters. In section 4 we describe the improved estimator that allows us
to obtain very accurate Monte Carlo data for the probability distribution of 
the staggered magnetization. In section 5 the results of numerical simulations
obtained with a loop-cluster algorithm are compared with the effective field
theory predictions. Finally, section 6 contains our conclusions.

\section{Effective Field Theory Predictions}

In this section we summarize the predictions of G\"ockeler and Leutwyler 
\cite{Goe91,Goe91a} which are derived from a 3-d $O(3)$-symmetric effective 
field theory. Although these authors had applications to the 3-d classical 
Heisenberg model in mind, their results also apply to the antiferromagnetic 
$(2+1)$-d quantum Heisenberg model. This microscopic model is defined by the 
Hamiltonian
\begin{equation}
H = J \sum_{\langle xy \rangle} \vec S_x \cdot \vec S_y -
\vec M_s \cdot \vec B_s,
\end{equation}
where $x$ and $y$ are nearest-neighbor sites on a square lattice with spacing
$a$. The spin $\tfrac{1}{2}$ operators $\vec S_x$ obey the standard commutation
relations 
\begin{equation}
[S_x^a,S_y^b] = i \delta_{xy} \varepsilon_{abc} S_x^c.
\end{equation}
Note that we work in natural units in which $\hbar = 1$. Furthermore, 
$\vec B_s$ represents an external staggered magnetic field and
\begin{equation}
\vec M_s = \sum_x (-1)^{(x_1+x_2)/a} \vec S_x
\end{equation}
is the staggered magnetization order parameter. In the infinite volume limit at
zero temperature and with $\vec B_s = 0$, $\vec M_s$ develops a non-zero
vacuum expectation value signaling the spontaneous breakdown of the $SU(2)_s$
spin symmetry down to its $U(1)_s$ subgroup.

The corresponding low-energy effective field theory is formulated in terms of 
the staggered magnetization order parameter field
\begin{equation}
\vec e(x) = (e_1(x),e_2(x),e_3(x)) \in S^2, \quad \vec e(x)^2 = 1,
\end{equation}
where $x = (x_1,x_2,t)$ is a point in Euclidean space-time. The leading terms 
in the effective action for the staggered magnetization field take the form
\begin{equation}
\label{action}
S[\vec e] = \int d^2x \ dt \ \left[\frac{\rho_s}{2}
\left(\p_i \vec e \cdot \p_i \vec e + 
\frac{1}{c^2} \p_t \vec e \cdot \p_t \vec e \right) - 
{\cal M}_s \vec e \cdot \vec B_s \right],
\end{equation}
where $\rho_s$ is the spin stiffness, $c$ is the spinwave velocity, and
${\cal M}_s$ is the staggered magnetization density. The corresponding 
partition function is given by
\begin{equation}
Z = \int D\vec e \ \exp(- S[\vec e]).
\end{equation}
Antiferromagnetic magnons have a ``relativistic'' dispersion relation with the 
spinwave velocity $c$ playing the role of the velocity of light. In fact, by
introducing $x_3 = c t$, the effective action can be written in the manifestly
Euclidean rotation-invariant form
\begin{equation}
S[\vec e] = \int d^3x \ \frac{1}{c} \left( \frac{\rho_s}{2}
\p_\mu \vec e \cdot \p_\mu \vec e - {\cal M}_s \vec e \cdot \vec B_s \right),
\end{equation}
which ensures Lorentz-invariance after analytic continuation from Euclidean to 
real time. It should be pointed out, however, that Euclidean rotation 
invariance is just an accidental symmetry of the leading terms of the effective
action. Since the underlying quantum Heisenberg model itself does not have this
symmetry, some of the higher-order four-derivative terms in the effective 
action break Euclidean rotation-invariance down to the discrete rotation 
subgroup of the square lattice. 

It is interesting to note that the ferromagnetic quantum Heisenberg model 
(which differs from the antiferromagnet only by the sign of the Hamiltonian) 
has very different symmetry properties at low energies. Unlike quantum 
antiferromagnets, quantum ferromagnets have a conserved order parameter --- the
uniform magnetization. Consequently, ferromagnetic magnons have a 
nonrelativistic dispersion relation and the corresponding effective action 
contains an additional Wess-Zumino term \cite{Leu94}, which breaks rotation
invariance between space and Euclidean time already at leading order. The 
resulting effective field theory for ferromagnetic magnons has been studied in 
detail in \cite{Hof99}. 

Here we concentrate entirely on antiferromagnets which are described by the 
effective action of eq.(\ref{action}). We consider the system in a periodic
cubic space-time volume $L \times L \times \beta$ with the inverse temperature
fixed at $\beta = L/c$ and with $\vec B_s = 0$. The space-time average of the 
staggered magnetization is given by
\begin{equation}
\vec \Phi = \frac{1}{2} \frac{1}{L^2 \beta} \int d^2x \ dt \ \vec e(x) =
\frac{1}{2} \frac{1}{L^3} \int d^3x \ \vec e(x).
\end{equation}
In contrast to \cite{Goe91,Goe91a}, we have included a factor $\tfrac{1}{2}$ in
the definition of $\vec \Phi$ because the quantum spins of the underlying 
Heisenberg model have $S = \tfrac{1}{2}$ while the effective field $\vec e(x)$ 
is normalized to 1. Due to the $O(3)$ symmetry, the probability distribution
\begin{equation}
p(\Phi) = \frac{1}{Z} \int D\vec e \ \exp(- S[\vec e]) \ 
\delta\left(\vec \Phi - \frac{1}{2} \frac{1}{L^3} \int d^3x \ \vec e(x)\right)
\end{equation}
of the mean staggered magnetization vector $\vec \Phi$ only depends on the 
magnitude $\Phi = |\vec \Phi|$. The distribution is normalized by
\begin{equation}
\label{norm}
4 \pi \int_0^\infty d\Phi \ \Phi^2 \ p(\Phi) = 1.
\end{equation}
The constraint effective potential $u(\Phi)$ represents the free energy density
of configurations constrained to a fixed mean staggered magnetization $\Phi$
and is determined by
\begin{equation}
p(\Phi) = {\cal N} \exp(- L^3 u(\Phi)),
\end{equation}
with the normalization derived in \cite{Goe91} given by
\begin{equation}
\label{normalization}
{\cal N} = \frac{1}{\widetilde{\cal M}^3_s} \ 
\frac{\rho_s L}{8 \pi^2 c e^{\beta_0}} \ 
\left[1 - \frac{c}{\rho_s L} \beta_1 + {\cal O}\left(\frac{1}{L^2}\right)
\right].
\end{equation}
Here $\widetilde{\cal M}_s = {\cal M}_s a^2$ is the staggered magnetization
per spin, while $\beta_0$ and $\beta_1$ are shape-coefficients of the 
space-time box. For the exactly cubic space-time volume considered here
$\beta_0 = 1.45385$ and $\beta_1 = 0.225785$. In the infinite volume and zero 
temperature limit the constraint effective potential approaches the infinite 
volume effective potential which is known to be a convex function of $\Phi$ 
\cite{Rai86}. In a finite volume, on the other hand, $u(\Phi)$ is not convex. 
An extensive variant of the intensive quantity $u(\Phi)$ is
\begin{equation}
U(\Phi) = L^3 u(\Phi).
\end{equation}

G\"ockeler and Leutwyler have used chiral perturbation theory to systematically
work out the finite-size effects of the constraint effective potential near its
minimum \cite{Goe91,Goe91a}
\begin{equation}
\label{UPhi}
U(\Phi) = U_0(\psi) + \frac{c}{\rho_s L} \ U_1(\psi) + 
{\cal O}\left(\frac{1}{L^2}\right).
\end{equation}
Here $U_0(\psi)$ and $U_1(\psi)$ are $L$-independent functions of the rescaled 
variable
\begin{equation}
\label{psi}
\psi = \frac{\rho_s L}{c} \ \frac{\Phi - \widetilde{\cal M}_s}
{\widetilde{\cal M}_s}.
\end{equation}
The leading order contribution to the constraint effective potential is given 
by an inverse Laplace transform
\begin{equation}
\label{U0}
\exp(- U_0(\psi)) = \int_{-\infty}^\infty dx \ \exp(- i x \psi + \Gamma(ix))
\end{equation}
of the function
\begin{equation}
\label{g0}
\Gamma(x) = \sum_{n=0}^\infty \frac{\beta_n x^n}{n!}. 
\end{equation}
Here the $\beta_n$ are shape-coefficients of the finite space-time box 
described in detail in appendix B of \cite{Has90}. As was pointed out in 
\cite{Goe91a}, the function $\Gamma(i x)$ is entirely kinematical and thus, 
unlike $U_1(\psi)$, the quantity $U_0(\psi)$ is universal, i.e.\ completely 
independent of the low-energy parameters. Consequently, $U_0(\psi)$ is the same
for all 3-d systems with an $O(3)$ symmetry spontaneously broken down to 
$O(2)$, including the 3-d classical and the $(2+1)$-d quantum Heisenberg model.
The $1/L$ correction to the leading contribution $U_0(\psi)$ is given by
\begin{eqnarray}
\label{U1}
U_1(\psi)&=&\psi + k_0 \exp(U_0(\psi)) \int_0^\infty dx \ x^2 \ \mbox{Re}
[\exp(- i x \psi + \Gamma(ix))] \nonumber \\
&=&\psi + \frac{k_0}{2} \left[\frac{d^2 U_0(\psi)}{d\psi^2} - 
\left(\frac{dU_0(\psi)}{d\psi}\right)^2\right].
\end{eqnarray}
Here $k_0$ is a low-energy constant related to the higher-order terms
\begin{equation}
\Delta S[\vec e] = - \int d^2x \ dt \
\left[h_1 (\vec e \cdot \vec B_s)^2 + h_2 \vec B_s^2\right]
\end{equation} 
in the effective action, which is given by
\begin{equation}
k_0 = \frac{2 \rho_s^3}{{\cal M}_s^2 c^2} (h_1 + h_2).
\end{equation}
It should be noted that the value of $k_0$ has no impact on 
eq.(\ref{normalization}) because it affects the normalization factor ${\cal N}$
only at higher orders in $1/L$.
 
Just as a non-zero quark mass in QCD explicitly breaks chiral symmetry and 
gives the pion its mass, a non-zero staggered magnetic field $\vec B_s$ 
explicitly breaks the $SU(2)_s$ spin symmetry and turns the magnons into 
pseudo-Nambu-Goldstone bosons with a non-zero mass $m$ determined at leading
order by
\begin{equation}
m^2 = \frac{{\cal M}_s B_s}{\rho_s c^2}, \quad B_s = |\vec B_s|.
\end{equation}
The constant $k_0$ also appears in the $B_s$-dependence of the field 
expectation value
\begin{equation}
\label{PhiBs}
|\langle \vec \Phi \rangle(B_s)| = \widetilde{\cal M}_s 
\left[1 + \frac{c}{\rho_s L}
\left(\sum_{n=0}^\infty \frac{\beta_{n+1}}{n!} (m c L)^{2n} - 
\frac{1}{(m c L)^2}\right)
+ k_0 \left(\frac{m c^2}{\rho_s}\right)^2 + {\cal O}(m^3)\right].
\end{equation}
It should be noted that eq.(\ref{PhiBs}) was derived in the $p$-regime of 
chiral perturbation theory in which $m c L \gg 1$ while $m c^2$, 
$c/L \ll \rho_s$. In particular, in eq.(\ref{PhiBs}) one cannot make $B_s$ (and
thus $m$) arbitrarily small, because one would otherwise enter the 
$\epsilon$-regime in which $m c L \approx 1$.

Besides the constraint effective potential, G\"ockeler and Leutwyler have also
derived analytic predictions for the first and second moment of the 
probability distribution $p(\Phi)$, including the 2-loop level. The resulting 
expressions are
\begin{eqnarray}
\label{moments}
&&\langle \Phi \rangle = 
\widetilde{\cal M}_s \left(1 + \frac{c}{\rho_s L} \beta_1 +
\frac{c^2}{\rho_s^2 L^2} \beta_2\right) + {\cal O}\left(\frac{1}{L^3}\right), 
\nonumber \\
&&\langle (\Phi - \langle \Phi\rangle)^2\rangle = 
\frac{\widetilde{\cal M}_s^2 c^2}{\rho_s^2 L^2} \beta_2 + 
{\cal O}\left(\frac{1}{L^3}\right).
\end{eqnarray}
For the cubic box considered here, the additional shape-coefficient is given 
by $\beta_2 = 0.010608$ \cite{Has90}.

Hasenfratz and Niedermayer have used the effective theory to derive the 
finite-size and finite-temperature effects of the staggered susceptibility
\begin{equation}
\label{chiscube}
\chi_s = \frac{{\cal M}_s^2 L^2 \beta}{3} \left\{1 + 2 \frac{c}{\rho_s L l} 
\beta_1(l) + \left(\frac{c}{\rho_s L l}\right)^2 \left[\beta_1(l)^2 
+ 3 \beta_2(l)\right] + {\cal O}\left(\frac{1}{L^3}\right) \right\}
\end{equation}
from a 2-loop calculation in the $\epsilon$-regime of magnon chiral
perturbation theory \cite{Has93}. Similarly, the uniform susceptibility takes 
the form
\begin{equation}
\label{chiucube}
\chi_u = \frac{2 \rho_s}{3 c^2} \left\{1 + \frac{1}{3} \frac{c}{\rho_s L l} 
\widetilde\beta_1(l) + \frac{1}{3} \left(\frac{c}{\rho_s L l}\right)^2
\left[\widetilde\beta_2(l) - \frac{1}{3} \widetilde\beta_1(l)^2 - 6 \psi(l)
\right] + {\cal O}\left(\frac{1}{L^3}\right) \right\}.
\end{equation}
Here $l = (\beta c /L )^{1/3}$ determines the shape of an approximately cubic 
space-time box of size $L \times L \times \beta$, with $\beta c \approx L$. The
functions $\beta_i(l)$, $\widetilde\beta_i(l)$, and $\psi(l)$ are known 
shape-coefficients \cite{Has90,Has93}. For an exactly cubical space-time volume
(i.e.\ for $l = 1$) the result of eq.(\ref{chiscube}) agrees with 
eq.(\ref{moments}) since
\begin{equation}
\label{susc}
\langle (\Phi - \langle \Phi\rangle)^2\rangle + \langle \Phi\rangle^2 = 
\langle \Phi^2 \rangle = \frac{3 \chi_s \widetilde{\cal M}^2_s}
{L^2 \beta {\cal M}^2_s} = \frac{3 \chi_s a^4}{L^2 \beta}.
\end{equation}
The factor 3 arises due to the three components of the staggered magnetization
vector.

\section{Determination of the Low-Energy Parameters ${\cal M}_s$, $\rho_s$, and $c$}

The susceptibilities $\chi_s$ and $\chi_u$ have been calculated numerically for
the antiferromagnetic spin $\tfrac{1}{2}$ quantum Heisenberg model on the 
square lattice using the very efficient loop-cluster algorithm \cite{Wie94}. By
comparing the Monte Carlo data with the effective theory predictions of 
eqs.(\ref{chiscube}) and (\ref{chiucube}), the low-energy parameters 
${\cal M}_s = 0.3074(4)/a^2$, $\rho_s = 0.186(4) J$, and $c = 1.68(1) J a$ have
been determined with high precision. At very low temperatures, one enters the 
cylindrical regime of space-time volumes with $\beta c \gg L$ in which the 
$\delta$-expansion of chiral perturbation theory applies. In this case, the 
staggered magnetization acts as a quantum mechanical rotor, again resulting in 
characteristic finite-volume effects \cite{Has93}. By simulating cylindrical 
space-time volumes using the continuous-time variant of the loop-cluster 
algorithm, and again comparing with the corresponding predictions of the 
low-energy effective theory, the values of the low-energy parameters, 
previously obtained from the cubical space-time regime, have been verified 
independently \cite{Bea96}. Using the value of the exact mass gap of the 2-d 
$O(3)$ model \cite{Has90b}, a result of Chakravarty, Halperin, and Nelson for 
the finite-temperature correlation length \cite{Cha89} was extended by 
Hasenfratz and Niedermayer \cite{Has90c} who obtained
\begin{equation}
\xi = \frac{e}{8} \ \frac{c}{2 \pi \rho_s} \exp(2 \pi \rho_s \beta)
\left[1 - \frac{1}{4 \pi \rho_s \beta} + 
{\cal O}\left(\frac{1}{\rho_s^2 \beta^2}\right)\right].
\end{equation}
This expression is valid in space-time volumes with a slab geometry, i.e.\ for
$L \gg \beta c$. While the data in the cubical and cylindrical regimes
determine the ratio $\rho_s/c^2$ with higher precision than $\rho_s$ and $c$
individually, Monte Carlo data for the very long correlation length in the slab
regime determine $\rho_s = 0.1800(5) J$ very precisely \cite{Bea98}. The 
combined analysis of all numerical data in cubical, cylindrical, and slab 
geometries resulted in $c = 1.657(2) J a$ and ${\cal M}_s = 0.30797(3)/a^2$ 
\cite{Bea98}. In a recent study using a zero-temperature valence-bond projector
method, Sandvik and Evertz obtained the very accurate result 
${\cal M}_s = 0.30743(1)/a^2$ \cite{San08}. Although the discrepancy between 
the two results for ${\cal M}_s$ is only in the permille range, it is 
statistically significant. In particular, it is important to clarify the
discrepancy because our present very accurate study is sensitive to such small
effects. For this purpose we have generated new data in the cubical regime for 
volumes ranging from $L/a = 16$ to 80. The largest volumes are substantially 
bigger than those of the original study \cite{Wie94}. In order to have an 
independent handle on the spinwave velocity $c$, in addition to the uniform 
susceptibility $\chi_u$, which is given by the temporal winding number $W_t$, 
we have also measured the spatial winding numbers $W_i$. The condition
\begin{equation}
\label{windings}
\langle W_t^2\rangle = \langle W_i^2\rangle
\end{equation}
determines an exactly cubical space-time box with $L = \beta c$. By varying
$\beta$ until eq.(\ref{windings}) is satisfied, we have determined 
$c = L/\beta = 1.6585(10) J a$ in excellent agreement with the result of 
\cite{Bea98}. By fitting the new data for $\chi_s$ and $\chi_u$ in the cubical 
regime to eqs.(\ref{chiscube}) and (\ref{chiucube}), we have obtained 
$\rho_s = 0.1808(4) J$ and ${\cal M}_s = 0.30744(3)/a^2$. The value for $\rho_s$
is again in excellent agreement with the previous results obtained in the 
cubical, cylindrical, and slab regimes. The new result for ${\cal M}_s$ agrees 
with the one of the original study in the cubical regime \cite{Wie94} and is
consistent with the result of \cite{San08}, which is about two permille lower 
than the result of \cite{Bea98}. We attribute this small but statistically 
significant discrepancy to an underestimation of the systematic errors of 
$\chi_s$ in the cylindrical regime data of \cite{Bea98}, related to the 
termination of the Seeley expansion described in \cite{Has93}. Until this issue
is completely clarified, we discard the cylindrical regime data and instead 
include the result of \cite{San08}. The best estimate of the low-energy
parameters obtained in cubic and slab geometries as well as at zero temperature
is then given by
\begin{equation}
\label{parameters}
{\cal M}_s = 0.30743(1)/a^2, \quad \rho_s = 0.1808(4) J, \quad 
c = 1.6585(10) J a.
\end{equation}

\section{Improved Estimator for the Distribution of the Staggered Magnetization}

The loop-cluster algorithm \cite{Eve93,Wie94,Bea96} is a very efficient 
numerical tool that allows us to perform high-accuracy numerical simulations of
the quantum Heisenberg model.The cluster algorithm connects the spin variables
to closed loop-clusters, which are completely independent of one another. All 
spins in a given cluster are then flipped simultaneously with 50 percent 
probability. A given spin configuration containing $N$ clusters is just one 
member of a sub-ensemble of $2^N$ equally probable configurations. An improved 
estimator substantially increases the statistics by analytically averaging a 
given observable over all $2^N$ configurations in the sub-ensemble. For the
quantum Heisenberg model, improved estimators have been constructed previously 
for the staggered and uniform susceptibilities $\chi_s$ and $\chi_u$ as well as
for the energy density \cite{Wie94}. The improved estimator for the 
distribution of the staggered magnetization to be constructed here is similar 
to the improved estimator for the topological charge distribution in the 
meron-cluster algorithm for the 2-d classical $O(3)$ model \cite{Bie95}, which 
has been combined with a re-weighting technique \cite{Wie89}.

In the loop-cluster algorithm, every cluster contributes additively to the 
total 3-component of the staggered magnetization. While it is straightforward 
to implement the improved estimator in continuous Euclidean time, it is most
easily explained in the discrete-time variant of the loop-cluster algorithm 
\cite{Eve93,Wie94}. In that case, the cluster size $|{\cal C}|$ (i.e.\ the
number of lattice points in a given cluster) determines the 3-component of the 
staggered magnetization of the cluster ${\cal C}$, which is proportional to 
$\pm |{\cal C}|$. Under cluster flip the staggered magnetization of a cluster 
changes sign. The distribution of the staggered magnetization is recorded as a 
histogram which is built iteratively using one cluster after the other. The 
initial histogram $p_1(m)$ (with $m \in \{-M,-M+1,...,0,...,M-1,M\}$, where $M$ 
is the number of space-time lattice points) is constructed from the first 
cluster as
\begin{equation}
p_1(m) = \frac{1}{2} \left[\delta_{m,|{\cal C}_1|} + 
\delta_{m,- |{\cal C}_1|}\right].
\end{equation}
The two entries of the initial histogram correspond to the two possible 
orientations of the first cluster, each arising with probability 
$\tfrac{1}{2}$. In the $i$-th iteration step (with $i \in \{2,3,...,N\}$) a new
histogram $p_i(m)$ is built from the previous one as
\begin{equation}
p_i(m) = \frac{1}{2}\left[p_{i-1}(m + |{\cal C}_i|) +
p_{i-1}(m - |{\cal C}_i|)\right].
\end{equation}
After $N$ steps, all clusters have been incorporated and the final histogram is
given by $p_N(m)$. Examples of histograms $p_N(m)$ obtained for two individual 
spin configurations are shown in figure 1. The example in the left panel 
contains one cluster that is bigger than all the other clusters together. 
Hence, the region around $m = 0$ is not sampled. In addition, there are two
relatively large clusters that give rise to the multiple peaks in the
distribution. In the example shown in the right panel, on the other hand, there
are two clusters of similar size, such that the region around $m = 0$ is also 
sampled.
\begin{figure}
\begin{center}
\epsfig{file=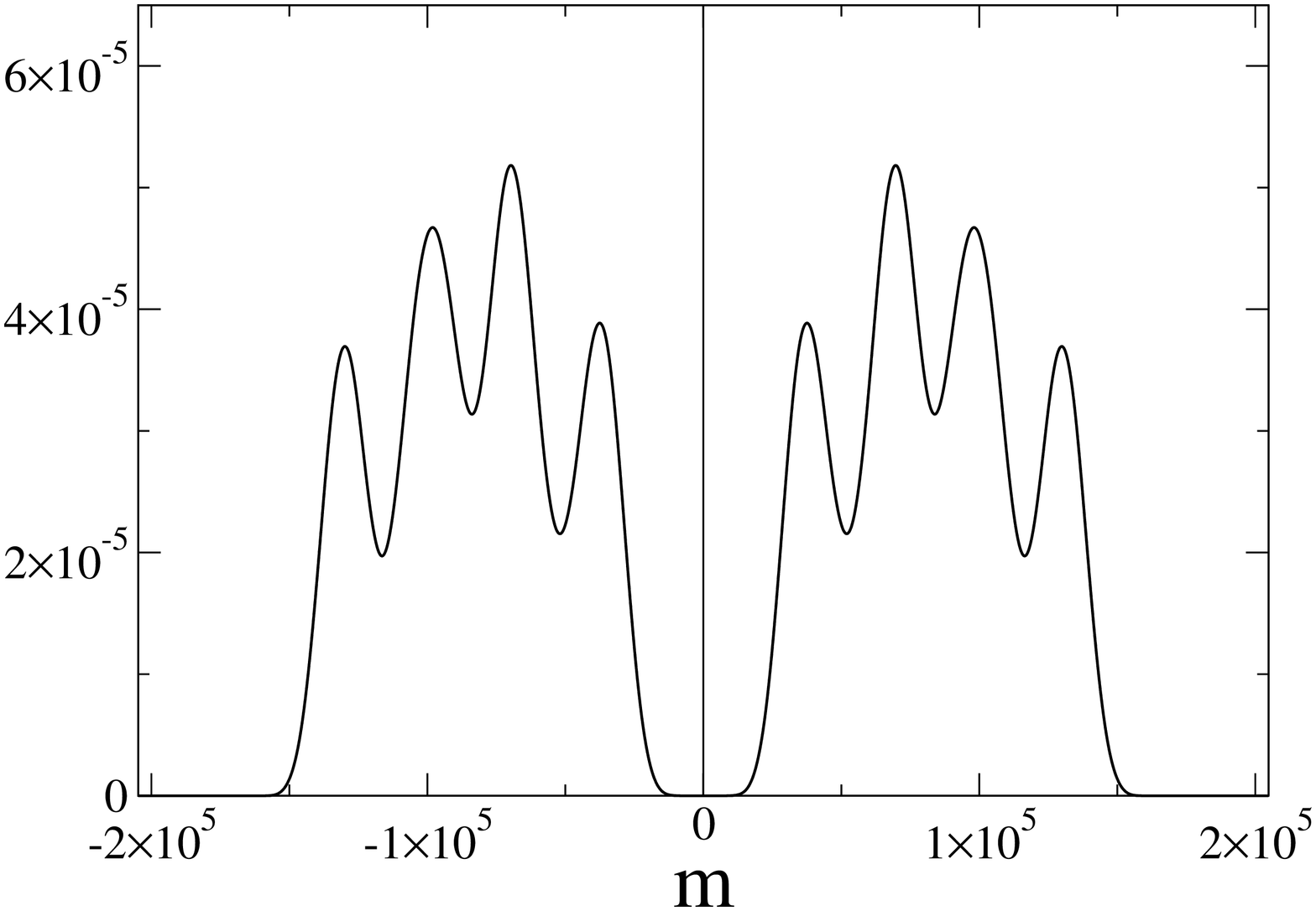,angle=-90,width=7cm}
\epsfig{file=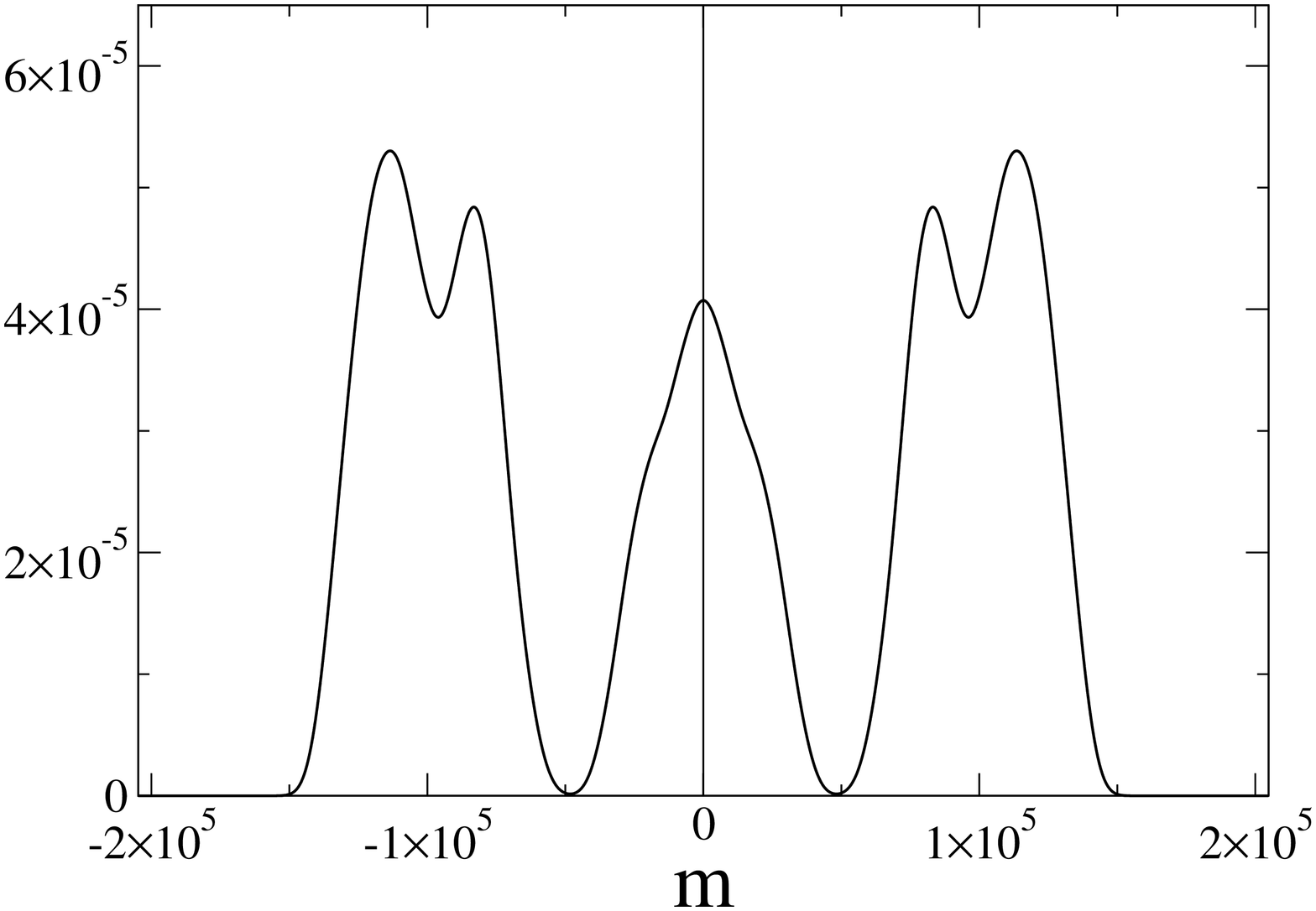,angle=-90,width=7cm}
\end{center}
\caption{\it Examples of histograms $p_N(m)$ obtained for two individual spin 
configurations on a $16^2$ lattice using the improved estimator.}
\end{figure}
The final probability distribution of the 3-component of the staggered 
magnetization 
\begin{equation}
p(m) = \langle p_N(m) \rangle,
\end{equation}
is the average of the histograms $p_N(m)$ for all configurations in the Markov 
chain generated by the cluster algorithm. By construction, this distribution is
properly normalized as
\begin{equation}
\sum_{m = -M}^M p(m) = 1.
\end{equation}

It should be noted that the numerical effort to build the improved estimator
is proportional to the number of lattice points $M$ and, in addition, 
proportional to the number of clusters. Since the number of clusters is 
proportional to the volume, the evaluation of the improved estimator requires a
computational effort proportional to $M^2$, and thus becomes rather 
time-consuming for large volumes. This is in contrast to the improved 
estimators for the susceptibilities $\chi_s$ and $\chi_u$ which only require a 
computational effort proportional to $M$. Of course, the improved estimator 
increases the statistics by a factor of $2^N$ which is exponential in the
volume. Hence, investing a polynomial effort $M^2$ should still be justified.
However, one should not forget that the $2^N$ configurations in a sub-ensemble
are not statistically independent. As we will see, the improved estimator works
very well and by far outperforms calculations done without it.

The mean value of the 3-component of the staggered magnetization $\Phi_3$ 
corresponding to a given value of $m$ is
\begin{equation}
\Phi_3 = \frac{m}{2 M}.
\end{equation}
The factor 2 in the denominator arises because we are dealing with quantum 
spins $\frac{1}{2}$. Hence, one can identify the probability distribution of 
the 3-component of the mean staggered magnetization as
\begin{equation}
\widetilde p(\Phi_3) \ d\Phi_3 = p(m).
\end{equation}
Due to the time-discretization chosen in \cite{Wie94} (which has four Trotter 
steps), all cluster sizes $|{\cal C}|$ are multiples of 8. Consequently, the 
non-zero entries of the histogram $p(m)$ correspond to values of $m$ which 
are also multiples of 8. This implies that
\begin{equation}
d\Phi_3 = \frac{4}{M}.
\end{equation}
Altogether, we thus obtain
\begin{equation}
\widetilde p(\Phi_3) = \frac{p(m)}{d\Phi_3} = 
\frac{M}{4} p(m), \quad \Phi_3 = \frac{m}{2M},
\end{equation}
with $m$ restricted to multiples of 8. By construction, in the Euclidean time
continuum limit the resulting probability distribution is normalized as
\begin{equation}
\int_{-\infty}^\infty d\Phi_3 \ \widetilde p(\Phi_3) = 1.
\end{equation}

We have simulated the spin $\tfrac{1}{2}$ quantum Heisenberg model on a square 
lattice with $L/a$ between $8$ and $24$ at inverse temperatures 
$\beta = L/c$ using the loop-cluster algorithm in its discrete-time variant 
\cite{Eve93,Wie94}. We have worked at a sufficiently small lattice spacing in 
Euclidean time, such that the systematic discretization error is negligible 
compared to the statistical errors. The probability distribution $p(\Phi_3)$ 
of the 3-component of the mean staggered magnetization $\Phi_3$ has been 
obtained using the improved estimator described above. A typical distribution
is shown in figure 2.
\begin{figure}
\begin{center}
\epsfig{file=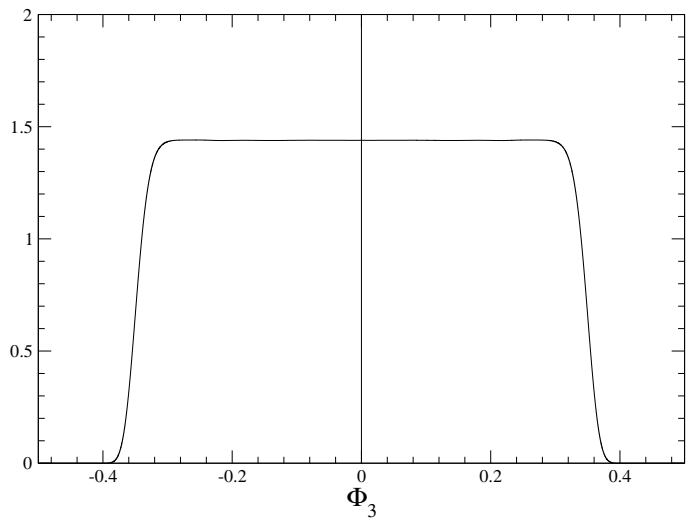,angle=-90,width=11cm}
\end{center}
\caption{\it Probability distribution $\widetilde p(\Phi_3)$ of the 3-component
of the staggered magnetization $\Phi_3$ on a $16^2$ lattice obtained with the 
improved estimator. The error bars of the distribution are on the order of the 
line width in this figure.}
\end{figure}
As we will see below, the information about the vicinity of the minimum of the 
constraint effective potential $u(\Phi)$ is contained in the region of $\Phi_3$
where $\widetilde p(\Phi_3)$ changes rapidly. Figures 2 and 3 compare Monte 
Carlo data obtained with and without the improved estimator, investing the same
amount of computer time in both cases. The error reduction of the improved 
estimator is very substantial.
\begin{figure}
\begin{center}
\epsfig{file=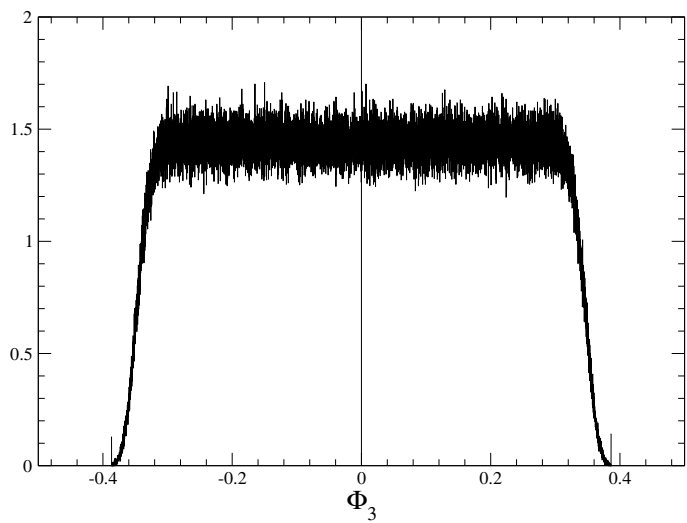,angle=-90,width=11cm}
\end{center}
\caption{\it The probability distribution $\widetilde p(\Phi_3)$ of the 
3-component of the staggered magnetization $\Phi_3$ on a $16^2$ lattice 
obtained without the improved estimator. Unlike in figure 2, which contains 
more than $4 \times 10^5$ bins, here the data have been compressed into $10^4$ 
bins, which substantially reduces their otherwise even larger variance.}
\end{figure}
By employing the re-weighting technique used in \cite{Wie89,Bie95}, one could
concentrate the statistics in the relevant region in which $p(\Phi_3)$ changes
rapidly. This should further improve the efficiency of our numerical method.

Due to the $O(3)$ symmetry, for a fixed magnitude $\Phi$ of the mean staggered 
magnetization vector $\vec \Phi$, its 3-component has a flat distribution
given by
\begin{equation}
\widetilde p_\Phi(\Phi_3) = \frac{1}{2 \Phi} \ \Theta_\Phi(\Phi_3).
\end{equation}
Here $\Theta_\Phi(\Phi_3)$ is a step function which is equal to 1 for
$\Phi_3 \in [-\Phi,\Phi]$ and zero otherwise. The probability distribution of 
the 3-component $\widetilde p(\Phi_3)$ and the one of the magnitude $p(\Phi)$ 
are related by
\begin{equation}
\widetilde p(\Phi_3) = 4 \pi \int_0^\infty d\Phi \ \Phi^2 \ p(\Phi) 
\ \widetilde p_\Phi(\Phi_3) = 
2 \pi \int_0^\infty d\Phi \ \Phi \ p(\Phi) \ \Theta_\Phi(\Phi_3),
\end{equation}
such that
\begin{eqnarray}
\frac{d\widetilde p(\Phi_3)}{d\Phi_3}&=&
2 \pi \int_0^\infty d\Phi \ \Phi \ p(\Phi) \ 
\frac{d\Theta_\Phi(\Phi_3)}{d\Phi_3} \nonumber \\
&=&- 2 \pi \int_0^\infty d\Phi \ \Phi \ p(\Phi) \ \delta(\Phi_3 - \Phi) =
- 2 \pi \Phi_3 \ p(\Phi_3).
\end{eqnarray}
Hence, given the Monte Carlo data for $\widetilde p(\Phi_3)$, we can extract
the probability distribution of the magnitude of the staggered magnetization as
\begin{equation}
\label{distributions}
4 \pi \Phi^2 \ p(\Phi) = - 2 \Phi \ \frac{d\widetilde p(\Phi)}{d\Phi}.
\end{equation}
This indeed ensures the correct normalization of eq.(\ref{norm}) because
\begin{equation}
4 \pi \int_0^\infty d\Phi \ \Phi^2 \ p(\Phi) = 
- 2 \int_0^\infty d\Phi \ \Phi \ \frac{d\widetilde p(\Phi)}{d\Phi} =
2 \int_0^\infty d\Phi \ \widetilde p(\Phi) = 
\int_{-\infty}^\infty d\Phi_3 \ \widetilde p(\Phi_3) = 1.
\end{equation}
Using eq.(\ref{distributions}), we have determined the probability 
distributions $p(\Phi)$ from $\widetilde p(\Phi_3)$ obtained using the improved
estimator. Some results for $4 \pi^2 \Phi^2 p(\Phi)$ are shown in figure 4. 
\begin{figure}
\begin{center}
\epsfig{file=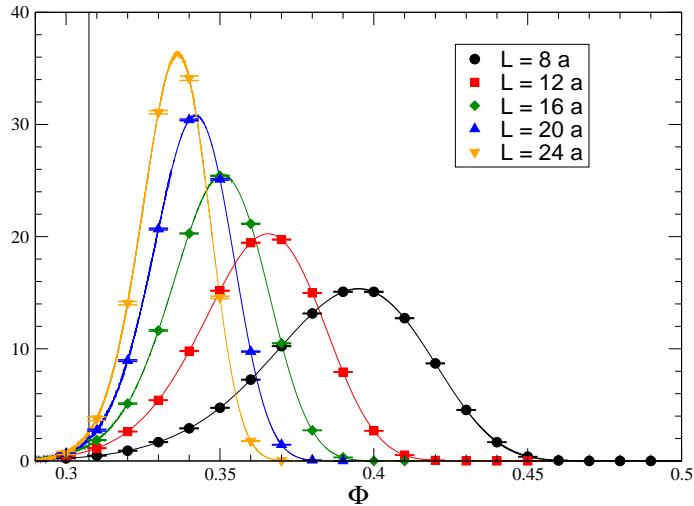,angle=-90,width=11cm}
\end{center}
\caption{\it Probability distributions $4 \pi \Phi^2 p(\Phi)$ of the magnitude 
of the staggered magnetization $\Phi$ for $L = 8a, 12a, 16a, 20a$, and $24a$. 
The vertical line at $\Phi = \widetilde{\cal M}_s = 0.30743(1)$ represents the 
$\delta$-function distribution of the infinite system.}
\end{figure}
As the volume increases the mean value of $\Phi$ decreases and the width of 
the distribution $p(\Phi)$ becomes narrower. It should be noted that the
distribution is not symmetric around its maximum. In the infinite volume limit,
the distribution turns into a $\delta$-function centered at 
$\Phi = \widetilde{\cal M}_s = 0.30743(1)$.

We also like to compute the first and second moment $\langle \Phi\rangle$ and
$\langle (\Phi - \langle \Phi\rangle)^2 \rangle$ of the distribution $p(\Phi)$.
Of course, this is trivial once we have computed $p(\Phi)$ using the improved
estimator which requires a computational effort proportional to $M^2$. If one
only wanted to compute the two moments but not $p(\Phi)$ itself, one may wonder
how to achieve this in the most efficient manner. As we noted before, the 
combination $\langle (\Phi - \langle \Phi\rangle)^2 \rangle + 
\langle \Phi\rangle^2 = \langle \Phi^2 \rangle$ is proportional to the 
staggered susceptibility which can be obtained using an improved estimator 
requiring a computational effort proportional to $M$. How can one determine
$\langle \Phi \rangle$ itself? Since $\Phi$ is the magnitude of the mean
staggered magnetization vector $\vec \Phi$, of which only the 3-component is
easily accessible in a quantum Monte Carlo simulation, this seems not entirely 
straightforward. However, using eq.(\ref{distributions}) one obtains
\begin{eqnarray}
\label{unimproved}
\langle \Phi \rangle&=&4 \pi \int_0^\infty d\Phi \ \Phi^3 \ p(\Phi) =
- 2 \int_0^\infty d\Phi \ \Phi^2 \ \frac{d\widetilde p(\Phi)}{d\Phi} =
4 \int_0^\infty d\Phi \ \Phi \ \widetilde p(\Phi) \nonumber \\
&=&2 \int_{-\infty}^\infty d\Phi_3 \ |\Phi_3| \ \widetilde p(\Phi_3) = 
2 \langle |\Phi_3| \rangle.
\end{eqnarray}
Hence, by just measuring $|\Phi_3|$ in a standard unimproved manner, one can 
determine $\langle \Phi \rangle$ with a computational effort proportional to 
$M$. The crucial question is how the statistical error achieved in this way 
compares with the one obtained by extracting $\langle \Phi\rangle$ from 
$p(\Phi)$ (whose construction requires a computational effort proportional to 
$M^2$). In order to investigate which of the two methods is more efficient, we 
have first determined $\langle \Phi\rangle$ from $p(\Phi)$ using the improved 
estimator described above. Then we have invested the same amount of computer 
time in an unimproved measurement of $2 \langle |\Phi_3| \rangle$. For 
$L/a = 20$ and $\beta = L/c$ the statistical error of $\langle \Phi \rangle$ 
obtained in this manner is a factor of 1.5 larger than when one uses the 
improved estimator. Hence, despite its computational effort proportional to 
$M^2$, thanks to the average over the large number $2^N$ of configurations in 
the sub-ensemble, the improved estimator is slightly better than the unimproved
method, even if one is only interested in the first moment 
$\langle \Phi \rangle$ and not in the distribution $p(\Phi)$ itself. 

In order to compare our Monte Carlo data with the expectation value
$|\langle \vec \Phi \rangle(B_s)|$ of eq.(\ref{PhiBs}), we also like to switch 
on a staggered magnetic field $\vec B_s = (0,0,B_s)$. The corresponding
probability distribution
\begin{equation}
\widetilde p(\Phi_3,B_s) = \frac{1}{Z(B_s)} \widetilde p(\Phi_3)
\exp\left(\frac{\Phi_3 B_s L^2 \beta}{a^2}\right).
\end{equation}
can be sampled using the loop-cluster algorithm with an additional Metropolis 
accept-reject step applied to each cluster flip, which takes into account the 
contribution $\exp(\Phi_3 B_s L^2 \beta/a^2)$ to the Boltzmann weight. Using an
ordinary unimproved estimator, one then simply measures
$|\langle \vec \Phi \rangle(B_s)| = |\langle \Phi_3 \rangle|$.

\section{Comparison of Monte Carlo Simulations and Effective Theory Predictions}

We have determined the first and second moment $\langle \Phi \rangle$ and 
$\langle (\Phi - \langle \Phi\rangle)^2\rangle$ of the distributions $p(\Phi)$,
which are compared with the effective field theory predictions of 
eq.(\ref{moments}) in table 1. The errors of the theoretical predictions 
result from the uncertainties in the low-energy parameters of 
eq.(\ref{parameters}). For the first moment the agreement is very good for 
$L/a \geq 10$. The absolute value of the second moment is very small and its 
statistical error is relatively large. Still, there are systematic 
discrepancies between the Monte Carlo data and the ${\cal O}(1/L^2)$ effective 
theory predictions of eq.(\ref{moments}). This discrepancy is well accounted 
for by additional ${\cal O}(1/L^3)$ corrections. Such corrections involve new 
low-energy parameters multiplying higher-order terms in the effective action. 
Their evaluation would require a 3-loop calculation which has not been worked 
out in the effective theory. Parameterizing the 3-loop terms with unknown 
coefficients $\alpha_1$ and $\alpha_2$, i.e.\
\begin{eqnarray}
&&\langle \Phi \rangle = 
\widetilde{\cal M}_s \left(1 + \frac{c}{\rho_s L} \beta_1 +
\frac{c^2}{\rho_s^2 L^2} \beta_2\right) + 
\alpha_1 \left(\frac{c}{\rho_s L}\right)^3 + 
{\cal O}\left(\frac{1}{L^4}\right), 
\nonumber \\
&&\langle (\Phi - \langle \Phi\rangle)^2\rangle = 
\frac{\widetilde{\cal M}_s^2 c^2}{\rho_s^2 L^2} \beta_2 + 
\alpha_2 \left(\frac{c}{\rho_s L}\right)^3 + 
{\cal O}\left(\frac{1}{L^4}\right), 
\end{eqnarray}
one obtains a good fit to the Monte Carlo data for $\alpha_1 = - 0.0017(2)$ and
$\alpha_2 = - 0.00042(2)$. One may conclude that precise calculations of the two
moments allow the determination of some combination of sub-leading low-energy 
parameters.
\begin{table}
\begin{center}
\begin{tabular}{|c|c|c|c|c|}
\hline
$L/a$ & $\langle \Phi \rangle_\text{MC}$ & $\langle \Phi \rangle_\text{theory}$
& $\langle (\Phi - \langle \Phi\rangle)^2\rangle_\text{MC}$ &
$\langle (\Phi - \langle \Phi\rangle)^2\rangle_\text{theory}$ \\
\hline
\hline
 8 & 0.38841(17) & 0.3913(2) & 
7.1(3) $\times 10^{-4}$ & 1.318(6) $\times 10^{-3}$ \\
\hline
10 & 0.37258(14) & 0.3738(2) & 
5.1(3) $\times 10^{-4}$ & 8.44(4) $\times 10^{-4}$ \\
\hline
12 & 0.36186(13) & 0.3624(1) & 
3.6(3) $\times 10^{-4}$ & 5.86(3) $\times 10^{-4}$ \\
\hline
14 & 0.35401(12) & 0.3543(1) & 
2.9(2) $\times 10^{-4}$ & 4.30(2) $\times 10^{-4}$ \\
\hline
16 & 0.34810(8) & 0.3483(1) & 
2.5(2) $\times 10^{-4}$ & 3.30(2) $\times 10^{-4}$ \\
\hline
18 & 0.34339(13) & 0.3437(1) & 
2.3(2) $\times 10^{-4}$ & 2.60(1) $\times 10^{-4}$ \\
\hline
20 & 0.33990(9) & 0.3400(1) & 
1.6(2) $\times 10^{-4}$ & 2.11(1) $\times 10^{-4}$ \\
\hline
22 & 0.33683(15) & 0.3369(1) &
1.6(2) $\times 10^{-4}$ & 1.74(1) $\times 10^{-4}$ \\
\hline
24 & 0.33462(19) & 0.3344(1) &
1.0(3) $\times 10^{-4}$ & 1.46(1) $\times 10^{-4}$ \\
\hline
\end{tabular}
\end{center}
\caption{\it Comparison of Monte Carlo data (MC) for the first and second 
moment $\langle \Phi \rangle$ and
$\langle (\Phi - \langle \Phi\rangle)^2\rangle$ of $p(\Phi)$ with the effective
theory predictions of eq.(\ref{moments}) at the 2-loop level. The errors of the
theory predictions are due to small uncertainties in the values of the 
low-energy parameters of eq.(\ref{parameters}). The discrepancy between the
Monte Carlo data and the effective field theory results is due to a 3-loop 
correction that was neglected in the theoretical predictions.}
\end{table}

Using $p(\Phi) = {\cal N} \exp(- L^3 u(\Phi))$, the probability distributions 
of figure 3 are readily converted into the corresponding constraint effective 
potentials $u(\Phi)$ shown in figure 5.
\begin{figure}
\begin{center}
\epsfig{file=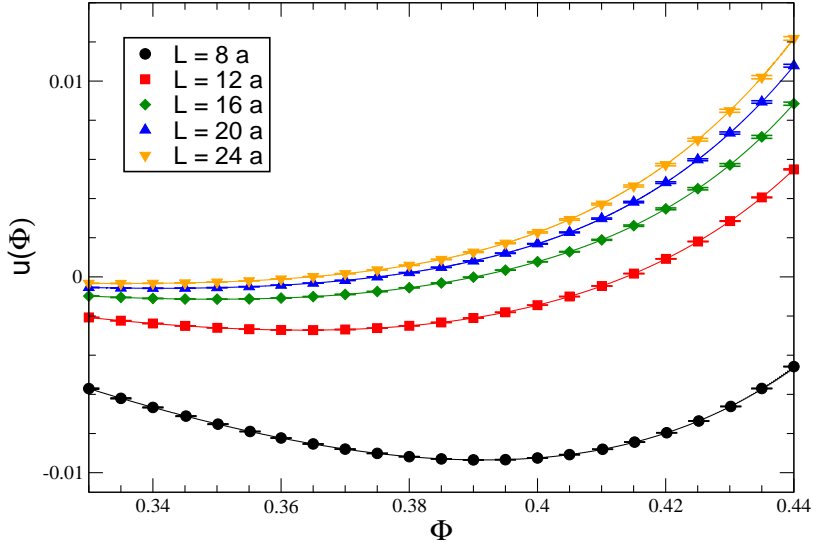,angle=-90,width=12cm}
\end{center}
\caption{\it Constraint effective potentials $u(\Phi)$ as functions of the 
magnitude of the staggered magnetization $\Phi$ for $L = 8a, 12a, 16a, 20a$, 
and $24a$. The constraint effective potential approaches a convex effective 
potential in the infinite volume limit.}
\end{figure}
With increasing volume the constraint effective potential approaches the convex
shape of the infinite volume effective potential. Using the rescaled variable 
$\psi = (\rho_s L/c) (\Phi - \widetilde{\cal M}_s)/\widetilde{\cal M}_s$, one 
can also consider the extensive quantity $U(\psi) = L^3 u(\Phi)$ which is shown
in figure 6.
\begin{figure}
\begin{center}
\epsfig{file=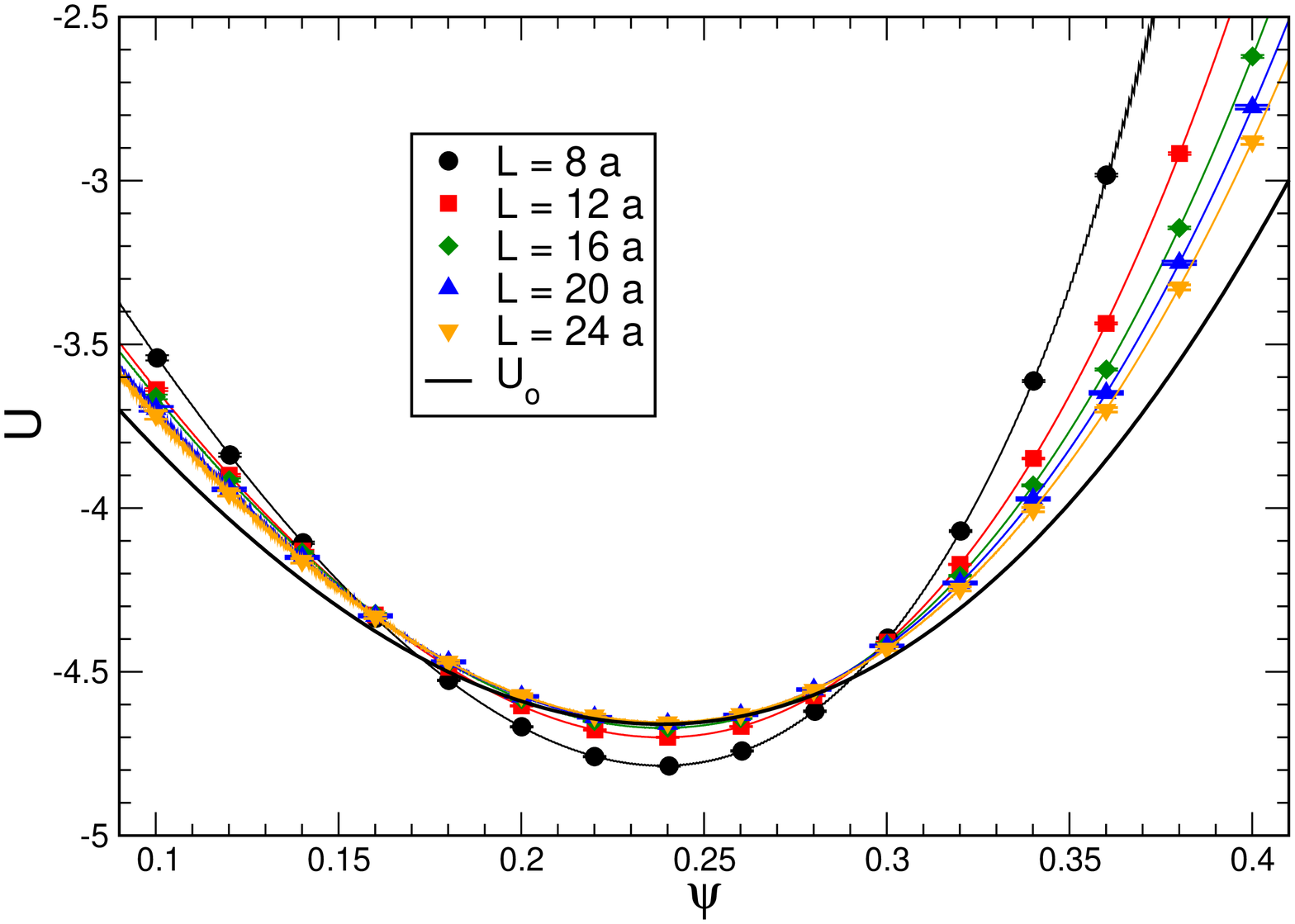,angle=-90,width=12cm}
\end{center}
\caption{\it The extensive quantity $U(\Phi)$ as a function of the rescaled 
variable 
$\psi = (\rho_s L/c) (\Phi - \widetilde{\cal M}_s)/\widetilde{\cal M}_s$ for 
$L = 8a, 12a, 16a, 20a$, and $24a$, compared to the analytic infinite volume 
result $U_0(\psi)$.}
\end{figure}
Expanding $U(\Phi) = U_0(\psi) + (c/\rho_s L) \ U_1(\psi) + {\cal O}(1/L^2)$, 
we have used the Monte Carlo data for $L/a$ between $12$ and $24$ to 
determine $U_0(\psi)$ and $U_1(\psi)$. Some values of the function $U_0(\psi)$ 
extracted from the numerical data are compared with the analytic result of 
eq.(\ref{U0}) in figure 7. It should be pointed out that the observed agreement
does not rely on any adjustable parameters. Even the normalization constant 
${\cal N}$ of eq.(\ref{normalization}), which fixes an additive constant in the
constraint effective potential, is predicted by the effective theory.
\begin{figure}
\vspace{1cm}
\begin{center}
\epsfig{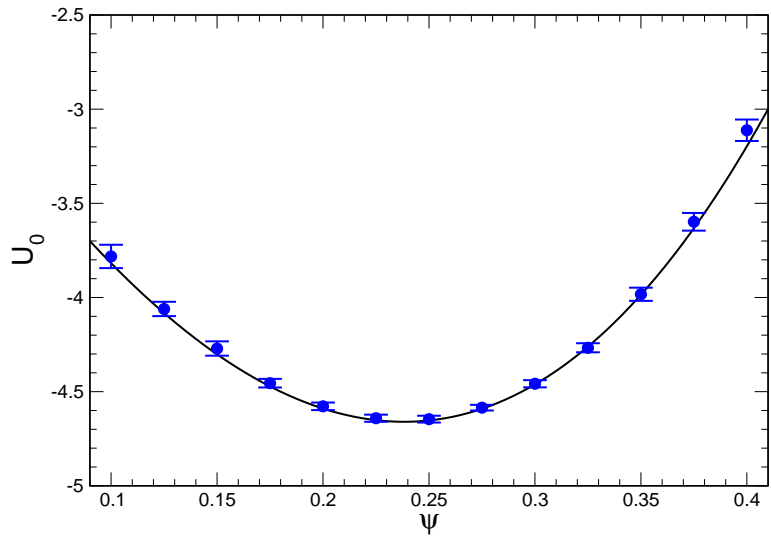}
\end{center}
\caption{\it The analytic result for the universal function $U_0(\psi)$ 
compared to numerical values obtained from a fit of the Monte Carlo data for
$U(\Phi)$ to eq.(\ref{UPhi}).} 
\end{figure}
As quantified in table 2, in the interval $\psi \in [0.10,0.35]$, i.e.\ around
the minimum of the constraint effective potential, the theoretical values and 
the simulation data for $U_0(\psi)$ agree within error bars.
\begin{table}
\begin{center}
\begin{tabular}{|c|c|c|c|c|}
\hline
$\psi$ & $U_0(\psi)_\text{MC}$ & $U_0(\psi)_\text{theory}$ \\
\hline
\hline
0.10 & - 3.782(62) & - 3.818 \\
\hline
0.15 & - 4.271(39) & - 4.302 \\
\hline
0.20 & - 4.578(22) & - 4.589 \\
\hline
0.25 & - 4.646(18) & - 4.653 \\
\hline
0.30 & - 4.458(20) & - 4.462 \\
\hline
0.35 & - 3.983(36) & - 3.986 \\
\hline
\end{tabular}
\end{center}
\caption{\it Comparison of Monte Carlo data (MC) for the universal function
$U_0(\psi)$ with the effective theory prediction of eqs.(\ref{U0}).}
\end{table}

In order to determine the low-energy parameter $k_0$, we have considered the
expectation value $|\langle \Phi \rangle(B_s)|$ as a function of the external 
staggered magnetic field given in eq.(\ref{PhiBs}). Since these results do not
require the probability distribution $p(\Phi)$, they could be obtained on 
larger volumes up to $L/a = 48$. The results summarized in table 3 give a good 
fit for
\begin{equation}
\label{k0}
k_0 = - 0.0037(3).
\end{equation}
\begin{table}
\begin{center}
\begin{tabular}{|c|c|c|c|}
\hline
$L/a$ & $B_s/J$ & $|\langle \vec \Phi \rangle(B_s)|$ \\
\hline
\hline
24 & 0.0300 & 0.35269(1) \\ 
\hline
24 & 0.0330 & 0.35472(1) \\ 
\hline
32 & 0.0175 & 0.34302(1) \\ 
\hline
32 & 0.0185 & 0.34396(1) \\ 
\hline
40 & 0.0100 & 0.33487(1) \\ 
\hline
40 & 0.0110 & 0.33615(1) \\ 
\hline
48 & 0.0075 & 0.33142(1) \\ 
\hline
48 & 0.0080 & 0.33219(1) \\ 
\hline
\end{tabular}
\end{center}
\caption{\it Monte Carlo data for $|\langle \vec \Phi \rangle(B_s)|$ which are
used in the determination of $k_0$.}
\end{table}

Using the theoretical prediction for $U_0(\psi)$, figure 8 compares values of 
the function $U_1(\psi)$ determined from the Monte Carlo data with the analytic
result of eq.(\ref{U1}), for the two values of $k_0$ at the edge of the
corresponding error band. In the interval $\psi \in [0.125,0.3]$ around the 
minimum of the constraint effective potential, the Monte Carlo data are 
consistent with the theoretical predictions of G\"ockeler and Leutwyler. We 
attribute the deviations outside this interval to effects of higher 
order. Indeed, as one sees in figure 6, the finite volume effects become larger
when $\psi$ moves away from the minimum of the constraint effective potential.
\begin{figure}
\begin{center}
\epsfig{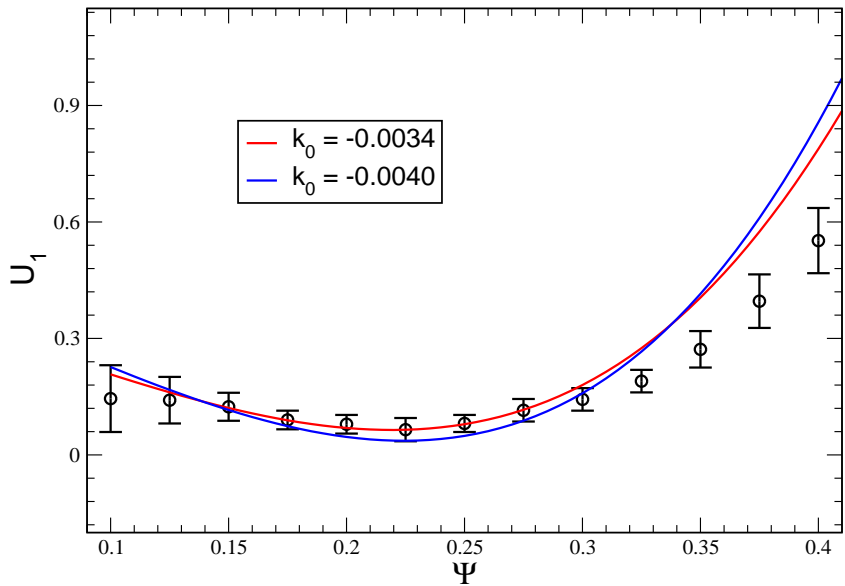}
\end{center}
\caption{\it The analytic result for the function $U_1(\psi)$ compared to 
numerical values obtained from a fit of the Monte Carlo data for $U(\Phi)$ to 
eq.(\ref{UPhi}), using the theoretical prediction of $U_0(\psi)$ as an input.}
\end{figure}

\section{Conclusions}

We have constructed an improved estimator for the probability distribution 
of the staggered magnetization in the quantum Heisenberg model. Using the 
improved estimator in a loop-cluster algorithm simulation, we have determined 
the first and second moment $\langle \Phi \rangle$ and
$\langle (\Phi - \langle \Phi\rangle)^2\rangle$ of the distribution $p(\Phi)$
of the magnitude $\Phi$ of the mean staggered magnetization vector $\vec \Phi$,
as well as the constraint effective potential $u(\Phi)$ (obtained from 
$p(\Phi) = {\cal N} \exp[- L^3 u(\Phi)]$) for different space-time volumes. The
Monte Carlo data are in excellent quantitative agreement with analytic 
predictions which G\"ockeler and Leutwyler derived from a systematic 
low-energy effective field theory. This demonstrates that the magnon effective
theory indeed provides correct predictions, order by order in a systematic
low-energy expansion. Thanks to the very efficient loop-cluster algorithm, the 
Heisenberg model is an excellent testing ground for the effective field theory 
method. Its quantitative success provides encouragement to also test the 
recently constructed systematic effective field theories for holes
\cite{Kae05,Bru06} and electrons \cite{Bru07} doped into an antiferromagnet
against numerical simulations. First results confirming the effective theory
have already been obtained for the $t$-$J$ model on the honeycomb lattice 
\cite{Jia08}. The results obtained in the study presented here should also be 
encouraging for lattice QCD simulations, were the numerical problem is much 
harder. Eventually, one may expect agreement between lattice QCD and chiral 
perturbation theory at the same level of accuracy as achieved in the condensed 
matter problem investigated in this paper.

\section*{Acknowledgments}

We have benefited from correspondence and discussions with B.\ B.\ Beard,
M.\ G\"ockeler, P.\ Hasenfratz, K.\ Jansen, F.\ Niedermayer, and 
H.\ Leutwyler. C.\ P.\ H.\
would like to thank the members of the Institute for Theoretical Physics at 
Bern University for their hospitality during a visit where this project was
initiated. The work of C.\ P.\ H.\ is supported by CONACYT Grant No.\ 50744-F
and by Grant Proyecto Cuerpo-Academico-56-UCOL. This work is supported in part 
by funds provided by the Schweizerischer Nationalfonds (SNF). The ``Center for 
Research and Education in Fundamental Physics'' at Bern University is supported
by the ``Innovations- und Kooperationsprojekt C-13'' of the Schweizerische 
Uni\-ver\-si\-t\"ats\-kon\-fe\-renz (SUK/CRUS).

\end{document}